\documentclass[aps,pra,twocolumn,showpacs,superscriptaddress]{revtex4}
\usepackage{graphicx}
\usepackage{amsmath}
\usepackage{amssymb}
\usepackage{lscape}
\usepackage{multirow}
\usepackage{longtable}

\input{epsf}

\begin{document}

\title{Analytical coupled-channels
treatment of two-body scattering in the presence of three-dimensional
isotropic spin-orbit coupling}

\author{Q. Guan}
\affiliation{Department of Physics and Astronomy,
Washington State University,
Pullman, Washington 99164-2814, USA}
\author{D. Blume}
\affiliation{Department of Physics and Astronomy,
Washington State University,
  Pullman, Washington 99164-2814, USA}

\date{\today}

\begin{abstract}
It is shown that the single-particle spin-orbit
coupling terms, 
which---in the cold atom context---are associated with synthetic gauge fields,
 can significantly and non-trivially modify the 
phase accumulation at small interparticle distances
even if the length scale $(k_{\text{so}})^{-1}$ 
associated with the spin-orbit coupling term
is significantly larger than the van der Waals length
$r_{\text{vdW}}$
that characterizes the two-body interaction potential.
A theoretical framework, which utilizes 
a generalized local frame transformation
and accounts for the phase accumulation analytically,
is developed. Comparison with numerical coupled-channels calculations
demonstrates that the phase accumulation can, to a
very good approximation, be described over a wide
range of energies by the free-space scattering phase
shifts---evaluated at a scattering energy that depends on $k_{\text{so}}$---and the
spin-orbit coupling strength $k_{\text{so}}$.
\end{abstract}

\maketitle

The tunability of
low-energy scattering parameters such as the $s$-wave
scattering length $a_s$ and $p$-wave scattering volume $V_p$ by
means of application of an external magnetic
field in the vicinity of a Feshbach resonance~\cite{chin} has
transformed the field of ultracold atom physics, providing
experimentalists with a knob to
``dial in'' the desired Hamiltonian.
This tunability 
has afforded the investigation 
of a host of new phenomena including the BEC-BCS crossover~\cite{block_rev,stringari_rev}.
Most theoretical treatments of these phenomena are formulated in
terms of a few scattering quantities such as $a_s$ and $V_p$,
which
properly describe the low-energy behavior of the two-body system.

The recent realization of spin-orbit coupled
cold atom systems~\cite{spielman} is considered another milestone, opening the
door for the observation of topological properties 
and 
providing a new platform with which to
study scenarios typically encountered in condensed matter systems
with unprecedented control~\cite{review_hui,review_coloquium,review_spielman}.
An assumption that underlies most theoretical
treatments of cold atom systems with synthetic gauge fields
is that the spin-orbit coupling term, i.e., the Raman laser that couples
the different internal states or the shaking of the lattice that 
couples different bands, leaves the atom-atom interactions 
``untouched''. More specifically, mean-field treatments ``simply''
add the single-particle spin-orbit coupling term to the
mean-field Hamiltonian and parameterize the atom-atom
interactions via  contact potentials with coupling strengths
that are calculated for the two-body van der Waals
potential without the spin-orbit
coupling terms~\cite{review_hui,review_hui2}.

Consistent with such 
mean-field approaches,
most two-body scattering studies derive 
observables based on the assumption that the 
two-body Bethe-Peierls
boundary condition, derived in the absence
of single-particle spin-orbit coupling terms, remains unaffected by the
spin-orbit coupling terms, provided an appropriate
``basis transformation'' is accounted for~\cite{pengzhang1,pengzhang2,xiaoling,chris,zhenhua1,guan,quasi-1D,gao}.
The underlying premise of these two-body and mean-field 
treatments is rooted
in scale separation, which
suggests that the free-space scattering length 
$a_s$ and scattering volume $V_p$ 
remain good quantities
provided $(k_{\text{so}})^{-1}$ is 
larger than the two-body van der Waals length $r_{\text{vdW}}$.
Indeed, model calculations for a square-well potential
in the presence of three-dimensional isotropic
spin-orbit coupling suggest that the above reasoning
holds, provided $1/a_s$ and $V_p$ are small~\cite{zhenhua2}.

This work revisits the question of how to obtain and parameterize
two-body scattering observables in the presence of
three-dimensional isotropic 
spin-orbit coupling.
Contrary to what has been reported in the literature,
our calculations for 
Lennard-Jones and square-well potentials show
that
the three-dimensional isotropic spin-orbit coupling
terms can impact the phase accumulation in the
small interparticle distance region
where the two-body interaction potential
cannot be neglected even if 
$(k_{\text{so}})^{-1}$ is notably larger
than $r_{\text{vdW}}$.
We observe non-perturbative changes of the 
scattering observables when $k_{\text{so}}$ changes by a small amount.
An analytical treatment, which reproduces the full
coupled-channels results such as the energy-dependent
two-body cross sections for the finite-range potentials
with high accuracy,
is developed.
Our analytical treatment relies, as do previous treatments~\cite{xiaoling,zhenhua1,zhenhua2,pengzhang1,pengzhang2,gao,guan,chris},
on separating the short- and large-distance regions.
The short-distance Hamiltonian is treated by
applying a gauge transformation,
followed by a rotation, that ``replaces''
the ${\mathbf{p}}$-dependent spin-orbit coupling term by an
${\mathbf{r}}$- and ${\mathbf{p}}$-independent diagonal
matrix (${\mathbf{r}}$ and ${\mathbf{p}}$ denote
the relative position and momentum vectors, respectively).
The diagonal terms, which can be interpreted as shifting the
scattering energy in each channel,
can introduce non-perturbative changes in the scattering observables for small
changes in $k_{\text{so}}$,
especially when $V_p$ is large.
We note that our derivation of the short-distance Hamiltonian,
although similar in spirit,
differs in subtle but
important ways from what is presented in Ref.~\cite{pengzhang1,pengzhang2}.

Our analytical framework also paves the way for designing
energy-dependent zero-range or $\delta$-shell
pseudo-potentials
applicable to systems with
spin-orbit coupling.
While energy-dependent pseudo-potentials have proven useful
in describing systems without
spin-orbit coupling~\cite{blume,julienne}, 
generalizations to systems with spin-orbit
coupling are non-trivial due to the more intricate nature of the
dispersion curves.
Our results suggest a paradigm shift in thinking about
spin-orbit coupled systems with non-vanishing two-body interactions.
While the usual approach is to assume that the short-distance behavior
or the effective coupling strengths are not impacted by the
spin-orbit coupling terms, our results suggest that they 
can be for specific
parameter combinations.
Even though our analysis is carried 
out for the case of 
three-dimensional isotropic 
spin-orbit coupling,
our results point toward a more general conclusion,
namely that
spin-orbit coupling terms
may, in general, notably modify the phase
accumulation in the short-distance region.

We consider two particles with position vectors
${\bf{r}}_j$ and masses $m_j$  ($j=1$ and $2$) 
interacting through a spherically symmetric
two-body potential ${V}_{\text{int}}(r)$
($r =| \mathbf{r}_1 - \mathbf{r}_2 |$). 
Both particles feel the isotropic  spin-orbit coupling 
term with strength $k_{\text{so}}$,
${V}_{\text{so}}^{(j)}=
\hbar k_{\text{so}}{{\bf{p}}}_j \cdot {{\pmb{\sigma}}}^{(j)} /m_j$,
where ${{\bf{p}}}_j$ denotes the canonical momentum operator of the 
$j$th particle and
${{\pmb{\sigma}}}^{(j)}$ the vector that contains the
three Pauli matrices ${\pmb{\sigma}}_x^{(j)}$,
${\pmb{\sigma}}_y^{(j)}$ and ${\pmb{\sigma}}_z^{(j)}$ for the $j$th particle.
Throughout, we assume that the 
expectation value of the total momentum
operator 
${{\bf{P}}}$
of the two-body system vanishes.
In this case, the total angular momentum operator
${{\bf{J}}}$,
${{\bf{J}}}={{\bf{l}}}+{{\bf{S}}}$,
of the two-particle system commutes with the system Hamiltonian
and the scattering solutions 
can be labeled by the quantum numbers $J$ and $M_J$; 
$M_J$ denotes
the projection quantum number,
${{\bf{l}}}$ is the relative orbital angular
momentum operator,
and ${{\bf{S}}}= \hbar ({\pmb{\sigma}}^{(1)}+{\pmb{\sigma}}^{(2)})/2$.

Separating off the center of mass degrees of freedom,
the relative Hamiltonian ${H}$ for the reduced mass $\mu$ particle
with relative momentum operator ${\mathbf{p}}$ can be written 
as a sum of the free-space Hamiltonian ${H}_{\text{fs}}$ and the spin-orbit
coupling term ${V}_{\text{so}}$,
${H}=
{H}_{\text{fs}} + {V}_{\text{so}}$,
where
\begin{eqnarray}
{H}_{\text{fs}} = \left[\frac{{\mathbf{p}}^2}{2\mu} +
{V}_{\text{int}}(r) \right] I_1 \otimes I_2
\end{eqnarray}
and
${V}_{so}= \hbar k_{so} {\pmb{\Sigma}} \cdot {\mathbf{p}} / \mu$
with ${\mathbf{\Sigma}} = (m_2 {\pmb{\sigma}}^{(1)} 
\otimes I_2 - m_1 I_1 \otimes {\pmb{\sigma}}^{(2)})/M$.
Here, $I_j$ denotes the $2 \times 2$ identity matrix that spans the 
spin degrees of freedom of the $j$th particle and $M$ the total mass,
$M=m_1+m_2$.
For each $(J,M_J)$ channel, 
the ${\bf{r}}$-dependent eigen functions
$\Psi^{(J,M_J)}$ are expanded as~\cite{gao,chris,guan}
\begin{eqnarray}
\Psi^{(J,M_J)}({\bf{r}})=\sum_{l,S} r^{-1} u_{l,S}^{(J)}(k, r) | J,M_J;l,S \rangle,
\end{eqnarray}
where the sum goes over 
$(l,S)=(0,0)$ and $(1,1)$ for $(J,M_J)=(0,0)$ and
over
$(l,S)=(J,0)$, $(J,1)$, $(J-1,1)$,
and $(J+1,1)$ for $J>0$.
In the $|J,M_J;l,S \rangle$ basis
(using the order of the states just given),
the scaled radial set of differential
equations for fixed $J$ and $M_J$ reads
$\underline{h}^{(J)} \underline{u}^{(J)} = E \underline{u}^{(J)}$,
where $\underline{h}^{(J)}$~\cite{footnote} denotes the scaled
radial Hamiltonian for a given $J$ (note
that the Hamiltonian is independent of the $M_J$ quantum number).
For $r>r_{\text{max}}$,
the interaction potential
$V_{\text{int}}$ can be neglected and
$\underline{u}^{(J)}$ is matched to the analytic asymptotic
$V_{\text{int}}=0$ solution~\cite{gao,chris,guan} 
\begin{eqnarray}
\label{eq_asym}
\underline{u}^{(J)} \xrightarrow[r> r_{\text{max}}] 
\, r \left( \underline{{\cal{J}}}^{(J)} - \underline{{\cal{N}}}^{(J)} \,
\underline{K}^{(J)} \right),
\end{eqnarray}
where $\underline{{\cal{J}}}^{(J)}$ and $\underline{{\cal{N}}}^{(J)}$ are 
matrices that contain the regular and irregular
solutions for finite $k_{\text{so}}$
(for $J=0$ and 1, explicit expressions are given in Ref.~\cite{guan}).
Defining the logarithmic derivative matrix $\underline{{\cal{L}}}^{(J)}(r)$
through $(\underline{u}^{(J)})' (\underline{u}^{(J)})^{-1}$,
where the prime denotes the partial derivative with respect to $r$,
the K-matrix is given by
\begin{eqnarray}
\label{eq_kmatrix}
\underline{{K}}^{(J)}=
\left[
\left(r {\underline{\cal{N}}}^{(J)} \right)' - 
{\underline{\cal{L}}}^{(J)}(r) 
\left(r {\underline{\cal{N}}}^{(J)} \right)'
\right] \times \nonumber \\
\left[
\left(r {\underline{\cal{J}}}^{(J)} \right)' - 
{\underline{\cal{L}}}^{(J)}(r) 
\left(r {\underline{\cal{J}}}^{(J)}
\right)
\right] \Big| _{r=r_{\text{max}}},
\end{eqnarray}
the S-matrix 
by
$\underline{S}^{(J)}=(\underline{I}+ \imath \underline{K}^{(J)})(\underline{I}- \imath 
\underline{K}^{(J)})^{-1}$, where $\underline{I}$ denotes the 
identity
matrix,
and the cross sections by
$\sigma_{\alpha \beta}= 
2 \pi |\underline{{{S}}}^{(J)}_{\beta \alpha}-\delta_{\alpha \beta}|^2/k_{\alpha}^2$, 
where  $\alpha$ and $\beta$ each take the 
values $1,2,\cdots$.

In general, the K-matrix has to be determined
numerically via coupled-channels calculations. In what follows,
we address the question whether $\underline{K}$ can, at least
approximately, be described in terms of the logarithmic 
derivative matrix
of the free-space Hamiltonian ${H}_{\text{fs}}$.
If the spin-orbit coupling term ${V}_{\text{so}}$
vanished in the small $r$ limit, one could straightforwardly
apply a projection or frame transformation approach~\cite{frame_blume,frame_greene,frame_fano,frame_harmin} 
that would 
project the inner small $r$ solution, calculated assuming that ${V}_{\text{so}}$ vanishes
in the inner region,
onto the outer large $r$ 
solution, calculated assuming that $V_{\text{int}}$ vanishes
in the outer region~\cite{suju_thesis}.
The fact that ${V}_{\text{so}}$ does not vanish in the small $r$ limit
requires, as we show below, 
a generalization of the frame transformation approach.

We start with the Hamiltonian $H$ and
define a new Hamiltonian $\tilde{H}$ through 
$T^{-1} H T$, where $T$ is an operator to be determined.
The solution $\tilde{\Psi}$ of the new Hamiltonian
is related to the solution $\Psi$ of $H$ through
$\tilde{\Psi} = T^{-1} \Psi$; here and in what follows we
drop the superscripts ``($J$, $M_J$)'' and ``($J$)" for notational convenience.
The operator $T$ reads $R U$, where 
$R= \exp(-\imath k_{\text{so}} {\mathbf{\Sigma}} \cdot {\mathbf{r}})$;
the form of $U$ is introduced later.
To calculate $H_R=R^{-1} H R$, we use
\begin{eqnarray}
\label{eq_rot1}
R^{-1}H_{\text{fs}} R =
H_{\text{fs}} - V_{\text{so}} - E_{\text{so}} 
\left[ {\mathbf{\Sigma}} \cdot {\mathbf{r}}, {\mathbf{\Sigma}} \cdot \mathbf{\nabla} \right]
+{\cal{O}}(\mathbf{r})
\end{eqnarray}
and
\begin{eqnarray}
\label{eq_rot2}
R^{-1} V_{\text{so}} R = V_{\text{so}} +
2 \left[ {\mathbf{\Sigma}} \cdot {\mathbf{r}}, {\mathbf{\Sigma}} \cdot \mathbf{\nabla} \right]
+{\cal{O}}(\mathbf{r}),
\end{eqnarray}
where $-\imath \hbar \mathbf{\nabla}=\mathbf{p}$ and $E_{\text{so}}=\hbar^2k_{\text{so}}^2/(2\mu)$
and where the notation ${\cal{O}}(\mathbf{r})$ indicates that
terms of order $r$ and higher are neglected 
($\mathbf{r}$ ``counts'' as being of order $r$ and $\mathbf{p}$
as being of order $r^{-1}$).
Adding Eqs.~(\ref{eq_rot1}) and (\ref{eq_rot2})
and neglecting the ${\cal{O}}(\mathbf{r})$ terms,
we find that the spin-orbit coupling term
$V_{\text{so}}$ is replaced by a commutator
that arises from the fact that the operator $\mathbf{\Sigma} \cdot \mathbf{p}$
does not commute with the exponent of $R$,
\begin{eqnarray}
\label{eq_sr1}
H_R^{\text{sr}} =
H_{\text{fs}} + E_{\text{so}} 
\left[ {\mathbf{\Sigma}} \cdot {\mathbf{r}}, {\mathbf{\Sigma}} \cdot \mathbf{\nabla} \right].
\end{eqnarray}
Here, the superscript ``sr'' indicates that this
Hamiltonian is only valid for small $r$.

Our goal is now to evaluate the second term on the
right hand side of Eq.~(\ref{eq_sr1}).
Defining the scaled short-distance Hamiltonian 
$h_R^{\text{sr}}$ through 
$r H_R^{\text{sr}} r^{-1}$
and expressing $h_R^{\text{sr}}$
in the $|J,M_J;l,S \rangle$ basis,
we find
\begin{eqnarray}
\underline{h}_R^{\text{sr}} =
\left( \frac{-\hbar^2}{2 \mu} \frac{\partial^2}{\partial r^2} + 
V_{\text{int}}(r) \right) I_1 \otimes I_2 + 
\underline{\cal{V}} + 
\underline{\epsilon},
\end{eqnarray}
where $\underline{\cal{V}}$
is a diagonal matrix with diagonal
elements $\hbar^2 l(l+1)/(2 \mu r^2)$.
For $J=0$, the matrix $\underline{\epsilon}$ is diagonal
with diagonal elements $-3E_{\text{so}}$
and $E_{\text{so}}$.
For $J>0$, in contrast, the 11 and 22 elements are,
in general, coupled:
\begin{eqnarray}
\underline{\epsilon} =E_{\text{so}}
\left(
\begin{array}{cccc}
-3 & c/M^2 & 0 & 0 \\
c/M^2 & -(\Delta M/M)^2 & 0 & 0 \\
0 & 0 & d_1/M^2 & 0 \\
0 & 0 & 0 & d_2/M^2  
\end{array}
\right),
\end{eqnarray}
where 
$\Delta M = m_1-m_2$,
$c=2 \sqrt{J(J+1)}(m_2^2 - m_1^2)$,
$d_1=-J M^2-(J+1)\Delta M^2$,
and $d_2=4m_1 m_2-d_1$.
Since the $r$-dependent 11 and 22 elements of $\underline{\cal{V}}$
are identical
(recall $l=J$ for these two elements), 
the matrix $\underline{U}$, which is defined 
such that 
$\underline{U}^{-1}  \underline{\epsilon} \underline{U}$
is diagonal,
also diagonalizes $\underline{h}_{R}^{\text{sr}}$,
i.e., the short-range
Hamiltonian
$\tilde{\underline{h}}_T^{\text{sr}}=
\underline{U}^{-1}  \underline{h}_R^{\text{sr}} \underline{U}$
is diagonal.
This implies that
the scaled radial short-distance Schr\"odinger
equation
$\tilde{\underline{h}}_T^{\text{sr}} \underline{v} = E \underline{v}$
can be solved using standard propagation schemes such as the Johnson
algorithm~\cite{johnson}.
This Schr\"odinger equation differs from
the ``normal"
free-space Schr\"odinger equation by channel-specific energy shifts.
These shifts
introduce a non-trivial modification of the phase accumulation
in the short-distance region
and---if a zero-range or $\delta$-shell
pseudo-potential description was
used---of the boundary condition. While the energy shifts do, in
many cases, have a negligible effect, our
analysis below shows that they can introduce
non-perturbative corrections in experimentally relevant parameter regimes.
The channel-specific energy shifts
are not taken into account in Ref.~\cite{pengzhang2}.

To relate the logarithmic derivative 
matrix 
${\underline{\tilde{\cal{L}}}}^{\text{sr}}(r) = 
 \underline{v}'
\underline{v}^{-1}$
of the scaled short-distance Hamiltonian
$\tilde{\underline{h}}_T^{\text{sr}}$ to the logarithmic 
derivative matrix ${\underline{{\cal{L}}}}(r)$,
the ``$T$-operation'' needs to be ``undone''. 
Assuming that the short-distance Hamiltonian
provides a faithful description, i.e., assuming that the higher-order
correction terms can, indeed, be neglected for $r< r_{\text{max}}$,
we obtain
\begin{eqnarray}
\label{eq_logder}
{\underline{{\cal{L}}}}(r_{\text{max}}) \approx
\left\{ \underline{T} {\underline{\tilde{\cal{L}}}}^{\text{sr}}(r)
\underline{T}^{-1} -
\underline{T}  \left(\underline{T}^{-1}
\right)'
\right\} \Big|_{r=r_{\text{max}}}.
\end{eqnarray}

To illustrate the results,
we focus on the $J=0$ subspace.
Denoting the usual free-space phase shifts 
at scattering energy $\hbar^2 k^2/ (2 \mu)$
for the interaction potential $V_{\text{int}}$ 
for the $s$-wave and $p$-wave channels 
by
$\delta_s(k)$ and $\delta_p(k)$,
respectively, the short-range K-matrix
$\tilde{\underline{K}}^{\text{sr}}$
for the Hamiltonian $\underline{\tilde{h}}^{\text{sr}}_T$
has the diagonal elements $\tan (\delta_s(k_s))$
and $\tan (\delta_p(k_p))$, where
$\hbar^2 k_s^2 /(2 \mu) = E+3 E_{\text{so}}$
and
$\hbar^2 k_p^2 /(2 \mu) = E- E_{\text{so}}$.
If we now, motivated by the concept of
scale separation, make the assumption that the phase
shifts $\tan (\delta_s(k_s))$
and $\tan (\delta_p(k_p))$ are accumulated
at $r=0$ and correspondingly take the $r_{\text{max}} \rightarrow 0$ limit of
Eq.~(\ref{eq_kmatrix}) with
$\underline{\cal{L}}^{(J)}$ given by the right hand
side of Eq.~(\ref{eq_logder}), we obtain the
following zero-range 
K-matrix,
\begin{widetext}
\begin{eqnarray}
\label{eq_zrmodel}
\underline{K}^{\text{zr}}
=-\frac{a_s(k_s)}{k_{+}-k_{-}}
\begin{bmatrix}
&k_{+}^2 &k_{+}k_{-} \\
&k_{+}k_{-} &k_{-}^2\\
\end{bmatrix}
-\frac{V_p(k_p)}{k_{+}-k_{-}}
\begin{bmatrix}
&k_{+}^2(k_{-}-k_{\text{so}})^2 &k_{+}k_{-}(k_{+}-k_{\text{so}})(k_{-}-k_{\text{so}}) \\
&k_{+}k_{-}(k_{+}-k_{\text{so}})(k_{-}-k_{\text{so}}) &k_{-}^2(k_{+}-k_{\text{so}})^2\\
\end{bmatrix},
\end{eqnarray}
\end{widetext}
where $\hbar k_{\pm}=\pm\sqrt{2\mu(E+E_{\text{so}})}-\hbar k_{\text{so}}$. 

\begin{figure}
\vspace*{0.6cm}
\hspace*{0cm}
\includegraphics[width=0.3\textwidth]{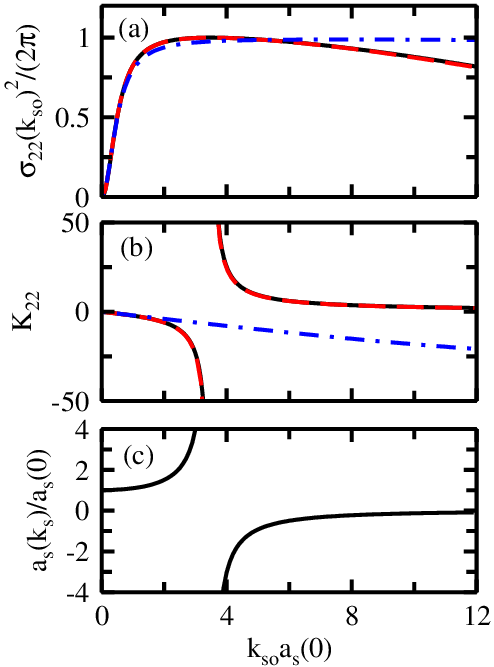}
\vspace*{-0.2cm}
\caption{
(Color online)
Large $a_s(0)$ case.
The black solid line shows 
(a) the scaled partial cross section $\sigma_{22}(k_{\text{so}})^2/(2 \pi)$
and
(b) the K-matrix element $K_{22}$
for $E=0$
as a function
of $k_{\text{so}} a_s(0)$ 
for the Lennard-Jones potential 
with $a_s(0)/r_{\text{vdW}}=24.42$ and  
$V_p(0)/(r_{\text{vdW}})^3=-0.2380$
(this potential supports two $s$-wave bound states in free space).
The
red dashed line 
shows the result for the zero-range model developed in this work
[see Eq.~(\ref{eq_zrmodel})];
the numerical results for the Lennard-Jones potential and the
model are indistinguishable on the scale shown. 
To illustrate the importance of the energy
shifts,
the blue dash-dotted line shows the results
for the zero-range model that artificially 
neglects the energy shifts.
The solid line in
(c) 
shows the scaled energy-dependent $s$-wave scattering length
$a_s(k_s)/a_s(0)$, where $\hbar^2 k_s^2=6 \mu E_{\text{so}}$. 
}
\label{fig1}
\end{figure}

\begin{figure}
\vspace*{0.6cm}
\hspace*{0cm}
\includegraphics[width=0.3\textwidth]{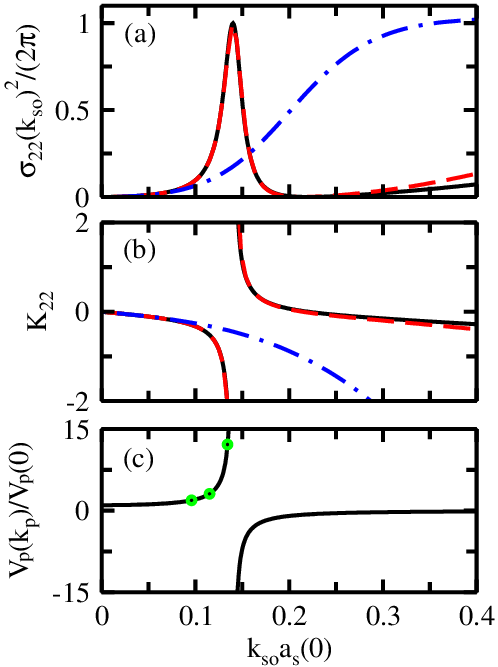}
\vspace*{-0.2cm}
\caption{
(Color online)
Large $V_p(0)$ case.
The black solid line shows
(a) the scaled partial cross section $\sigma_{22}(k_{\text{so}})^2/(2 \pi)$
and
(b) the K-matrix element $K_{22}$
for $E=0$
as a function
of $k_{\text{so}} a_s(0)$ 
for the Lennard-Jones potential 
with $a_s(0)/r_{\text{vdW}}=0.9591$ and  $V_p(0)/(r_{\text{vdW}})^3=26.61$,
corresponding to $a_s(0)/(V_p(0))^{1/3}=0.3213$
(this potential supports 4 four $s$-wave bound states in free space).
The
red dashed line 
shows the result for the zero-range model developed in this work
[see Eq.~(\ref{eq_zrmodel})];
the model reproduces the
numerical results 
excellently for $k_{\text{so}} a_s(0) \lesssim 0.3$.
The blue dash-dotted line shows the results
for the zero-range model that artificially 
neglects the energy shifts.
The solid line in
(c) 
shows the scaled energy-dependent $p$-wave scattering volume
$V_p(k_p)/V_p(0)$, where $\hbar^2 k_p^2 = -2 \mu E_{\text{so}}$.
The green circles mark three of the four $k_{\text{so}} a_s(0)$ values considered
in Fig.~\ref{fig5}.
}
\label{fig2}
\end{figure}

To validate our analytical results, we
perform numerical coupled-channels calculations.
Since
the wave function in the $J=0$ subspace
is anti-symmetric under the simultaneous
exchange of the spatial and spin
degrees of freedom of the two particles,
the solutions apply to two identical 
fermions. 
The Schr\"odinger equation for the  Lennard-Jones potential 
$V_{\text{LJ}}(r)=C_{12}/r^{12}-C_6/r^6$,
with $C_6$ and $C_{12}$ denoting positive coefficients,
is 
solved numerically~\cite{JCP_paper}.
The solid lines in Figs.~\ref{fig1} and \ref{fig2} show 
the partial cross section $\sigma_{22}$ and the K-matrix element $K_{22}$
as a function of $k_{\text{so}}$ for
vanishing scattering energy $E$
for a two-body potential with large $a_s(0)$ and
large $V_{p}(0)$, respectively.
The dashed lines show the
results predicted by our zero-range model that
accounts for the spin-orbit coupling induced energy shifts.
This model provides an excellent description
of the numerical results for the Lennard-Jones potential,
provided the length $(k_{\text{so}})^{-1}$
associated with the spin-orbit coupling term is
not too small compared to the van der Waals length
$r_{\text{vdW}}$,
where $r_{\text{vdW}}$ is given by $(2 \mu C_6/\hbar^2)^{1/4}$ 
(in Figs.~\ref{fig1} and \ref{fig2}, the largest
$k_{\text{so}} r_{\text{vdW}}$ considered corresponds to 
$0.4913$ and $0.4171$, respectively).

The dash-dotted lines in Figs.~\ref{fig1} and \ref{fig2}
show $\sigma_{22}$ and $K_{22}$ for the zero-range model when
we set the spin-orbit coupling induced energy shifts
artificially to zero.
In this case,
the divergence in the $K_{22}$ matrix element
at finite $k_{\text{so}}$ is not reproduced.
For large $a_s(0)$ [see Fig.~\ref{fig1}(a)], the
model without energy shifts introduces 
deviations at the few percent level in the cross section
$\sigma_{22}$.
For large $V_p(0)$ [see Fig.~\ref{fig2}(a)], in contrast,
the model without the energy shifts provides
a quantitatively and qualitatively poor
description of the cross section $\sigma_{22}$
even
for relatively small 
$k_{\text{so}}$
($k_{\text{so}} a_s(0) \gtrsim 0.05$).
Figures~\ref{fig1}(c) and \ref{fig2}(c)
demonstrate that the divergence of the $K_{22}$
matrix element occurs when the free-space scattering length
$a_s(k_s)$, calculated at energy $3E_{\text{so}}$, or
the free-space scattering volume
$V_p(k_p)$, calculated at energy $-E_{\text{so}}$, diverge.
We find that this occurs roughly
when 
$a_s(0) k_{\text{so}} \approx 10$
and
$(V_p(0))^{1/3}k_{\text{so}} \approx 0.21$;
we checked that this holds quite generally,
i.e., not only for the parameters considered in the figures.
In Figs.~\ref{fig1}(c) and \ref{fig2}(c),
the ``critical'' $k_{\text{so}}$ values correspond
to $k_{\text{so}} r_{\text{vdW}}=0.1423$
and 
$k_{\text{so}} r_{\text{vdW}}=0.1462$, respectively.
For comparison, using the $k_{\text{so}}$ value
for the one-dimensional realization of Ref.~\cite{spielman}
and assuming $r_{\text{vdW}}=100 a_0$,
one finds $k_{\text{so}} r_{\text{vdW}} \approx 0.03$. 
This suggests that the phenomena discussed in the context
of Figs.~\ref{fig1} and \ref{fig2} should be relevant to
future realizations of three-dimensional isotropic spin-orbit
coupling experiments.

\begin{figure}
\vspace*{0.6cm}
\hspace*{0cm}
\includegraphics[width=0.35\textwidth]{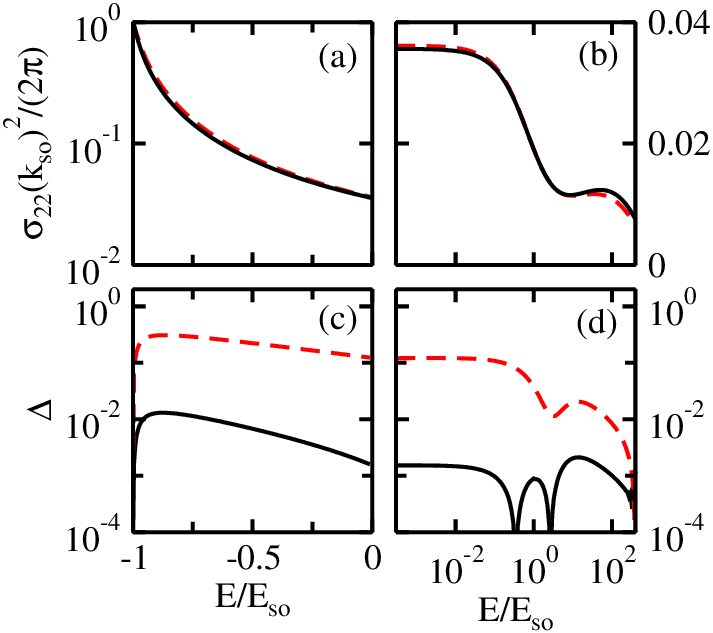}
\vspace*{-0.2cm}
\caption{
(Color online)
  Large $V_p(0)$ case.
  (a) and (b):
  The red dashed and black solid 
  lines 
show the scaled partial cross section
$\sigma_{22} (k_{\text{so}})^2/(2 \pi)$
for the 
Lennard-Jones and square-well potential,
respectively, 
as a function of the scattering energy $E$.
For both potentials, 
we have
$a_s(0)/(V_p(0))^{1/3}=0.3213$
[$V_p(0)>0$] and  $k_{\text{so}} a_s(0)=0.07673$.
The length scale associated with the
spin-orbit coupling is notably
larger
than the range of the potential
($k_{\text{so}} r_{\text{vdW}}=0.08$ for the
Lennard-Jones potential
and
$k_{\text{so}} r_{\text{sw}}=0.07676$ for the 
square-well potential).
(c) and (d):
The solid and dashed lines show the normalized difference $\Delta$
(see text) between the cross section for the Lennard-Jones potential and  the zero-range
model, obtained using Eq.~\eqref{eq_zrmodel}, and between that for the Lennard-Jones potential
and the zero-range model that neglects the
spin-orbit coupling induced energy shifts, respectively.
The zero-range model derived in this work (solid line) provides an
excellent description (the deviations are smaller than $1.3$~\% for the
data shown) over the entire energy regime.
Panels (a) and (c) cover negative $E$ 
(linear scale) while
panels (b) and (d) cover positive $E$
(logarithmic scale)].
}
\label{fig4}
\end{figure}

To further explore the two-particle 
scattering properties in the presence of spin-orbit coupling
for short-range potentials
with large free-space scattering volume $V_p(0)$,
Figs.~\ref{fig4}(a) and \ref{fig4}(b) show the partial
cross section $\sigma_{22}$ as a function of 
the scattering energy $-E_{\text{so}} \le E \le 0$
and $0 \le 0 \le 400E_{\text{so}}$, respectively,
for $a_s(0)/(V_p(0))^{1/3}=0.3213$
and $a_s(0) k_{\text{so}}=0.07673$. 
The results for the Lennard-Jones potential 
(dashed line) and
square-well potential (solid line)
are essentially indistinguishable on the
scale shown. 
To assess the accuracy of our zero-range model,
we focus on the Lennard-Jones potential and
compare the numerically determined partial cross section
$(\sigma_{22})^{\text{exact}}$ with the
partial cross section $(\sigma_{22})^{\text{zr}}$ predicted
using Eq.~(\ref{eq_zrmodel}).
Solid lines in Figs.~\ref{fig4}(c) and \ref{fig4}(d)
show the normalized difference $\Delta$,
defined through
$\Delta=
|(\sigma_{22})^{\text{zr}}-(\sigma_{22})^{\text{exact}}|/(\sigma_{22})^{\text{exact}}$.
The deviations are smaller than $1.3$\% for the scattering energies considered.
Neglecting the spin-orbit coupling
induced energy shifts in our zero-range model
and calculating the normalized difference,
we obtain the dashed lines
in Figs.~\ref{fig4}(c) and \ref{fig4}(d).
Clearly, the zero-range model
provides a faithful description
of the full
coupled-channels data
for the Lennard-Jones
potential only if the spin-orbit coupling induced energy shifts are included.

\begin{figure}
\vspace*{1.2cm}
\hspace*{0cm}
\includegraphics[width=0.35\textwidth]{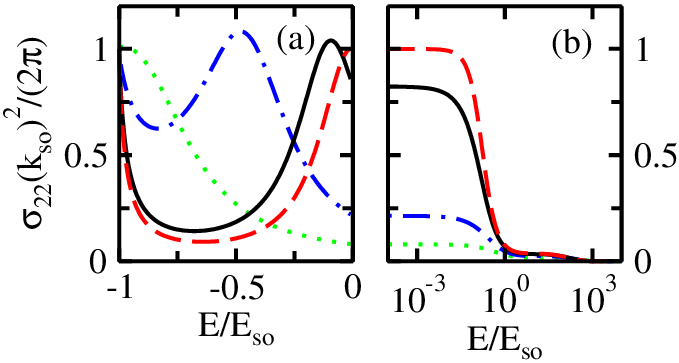}
\vspace*{-0.2cm}
\caption{
(Color online)
Scaled partial cross section
$\sigma_{22} (k_{\text{so}})^2/(2 \pi)$
for the Lennard-Jones potential
with
$a_s(0)/(V_p(0))^{1/3}=0.3213$ 
($V_p(0)>0$) and $a_s(0)/r_{\text{vdW}}=0.9591$
for four
different $k_{\text{so}}$
[the green dotted, blue
dash-dotted, black solid, and red dashed lines
correspond to
$k_{\text{so}} r_{\text{vdW}}=0.1$,
$k_{\text{so}} r_{\text{vdW}}=0.12$,
$k_{\text{so}} r_{\text{vdW}}=0.14$, and
$k_{\text{so}} r_{\text{vdW}}=0.146$, respectively]
as a function of the scattering energy $E$
[panel (a) covers negative $E$ 
(linear scale) while
panel (b) covers positive $E$
(logarithmic scale)].
}
\label{fig5}
\end{figure}

Figure~\ref{fig5} demonstrates that the non-quadratic single-particle dispersion relations have a profound impact on the low-energy 
scattering observables
for a large free-space scattering volume.
Specifically, the lines in Fig.~\ref{fig5} show the 
numerically obtained partial cross section
$\sigma_{22}$ as a function of the 
scattering energy for the same Lennard-Jones potential
as that used in Figs.~\ref{fig2} and~\ref{fig4} for four different
spin-orbit coupling strengths,
namely $k_{\text{so}} r_{\text{vdW}}=0.1$, $0.12$,
$0.14$ and $0.146$
[Fig.~\ref{fig4} used $k_{\text{so}} r_{\text{vdW}} =0.08$; three of the four $k_{\text{so}}$ values considered in Fig.~\ref{fig5} are marked by circles in Fig.~\ref{fig2}(c)].
Figure~\ref{fig5} shows that the 
partial cross section depends sensitively 
on the spin-orbit coupling strength
$k_{\text{so}}$. This can be understood by realizing
that a change in the spin-orbit coupling strength leads
to a significant change
of the $k_{\text{so}}$-dependent 
scattering volume $V_p(k_p)$.

This paper revisited 
two-body scattering in the presence of
single-particle interaction terms that
lead, in the absence of two-body interactions, to 
non-quadratic dispersion relations.
Restricting ourselves to three-dimensional
isotropic spin-orbit coupling terms and 
spin-independent central two-body interactions,
we developed an analytical coupled-channels
theory that connects the short- and large-distance
eigenfunctions using a generalized frame transformation.
A key, previously overlooked result of our treatment is
that the gauge transformation 
that converts the short-distance Hamiltonian 
to the ``usual form'' (i.e., a form
without linear momentum dependence)
introduces channel-dependent energy shifts. 
These energy shifts
were then shown to appreciably alter the low-energy scattering
observables, especially in the regime where the 
free-space scattering volume is large.
To illustrate this, the $(J,M_J)=(0,0)$ channel was
considered exemplarily.
Our framework provides the first complete 
analytical description that consistently accounts
for all partial wave channels.
Moreover, the first numerical
coupled-channels results for two-particle
Hamiltonian with realistic Lennard-Jones potentials
in the presence of spin-orbit coupling terms were presented.
The influence of the revised zero-range formulation put
forward in this paper on two- and few-body bound states 
and on mean-field and beyond mean-field studies will be
the topic of future publications.

\section{acknowledgement}
\label{acknowledgement}
Support by the National Science Foundation through
grant number  
PHY-1509892 is gratefully acknowledged.
The authors acknowledge hospitality of 
and support 
(National Science Foundation under Grant No. NSF PHY-1125915)
by the KITP.
We thank J. Jacob for providing us with a copy
of his
coupled-channels code.


\begin{thebibliography}{100}
\bibitem{footnote}
Here and in what follows, underlined quantities are matrices.
\end{thebibliography}
\end{document}